\documentclass[prl,showkeys,preprintnumbers,amsmath,amssymb,unsortedaddress,aps,preprint]{revtex4-2}

\usepackage{microtype}
\usepackage[hidelinks]{hyperref}
\usepackage{graphicx}
\usepackage{amssymb}
\usepackage{amsmath}
\usepackage{amsfonts,bm}
\usepackage{textcomp}
\usepackage{bm}
\usepackage[e]{esvect}

\graphicspath{ {./figures/} }

\let\quat=\boldsymbol
\renewcommand{\vec}[1]{{\protect\vv{\quat{#1}}}}
\newcommand{\vecsub}[2]{{\protect\vv*{\quat{#1}}{#2}}}
\newcommand{\conj}[1]{#1^\dag}
\newcommand{\conjtext}{dagger}
\newcommand{\conjbr}[1]{(#1)^\dag}

\newcommand{\conjwbs}[1]{\left[#1\right]^\dag}
\newcommand{\uqI}{\vec{i}}
\newcommand{\uqJ}{\vec{j}}
\newcommand{\uqK}{\vec{k}}
\newcommand{\quatcomp}[4]{#1+#2\uqI+#3\uqJ+#4\uqK}
\newcommand{\quatcompconj}[4]{#1-#2\uqI-#3\uqJ-#4\uqK}
\newcommand{\quatcompvec}[3]{#1\uqI+#2\uqJ+#3\uqK}

\newcommand{\uO}{\mathrm{O}}

\newcommand{\uY}{\mathrm{Y}}
\newcommand{\uZ}{\mathrm{Z}}
\newcommand{\ux}{\mathrm{X}}
\newcommand{\uy}{\mathrm{Y}}
\newcommand{\uz}{\mathrm{Z}}
\newcommand{\uMs}{M_{\mathrm{S}}}
\newcommand{\umx}{m_{\mathrm{X}}}
\newcommand{\umy}{m_{\mathrm{Y}}}
\newcommand{\umz}{m_{\mathrm{Z}}}

\newcommand{\uvr}{\vec{r}}
\newcommand{\uvri}{\vecsub{r}{i}}
\newcommand{\uFH}{{\bm{\mathcal{H}}}}
\newcommand{\uFC}{{\bm{\mathcal{C}}}}
\newcommand{\uFB}{{\vec{\mathcal{B}}}}
\newcommand{\uFR}{{\bm{\mathcal{R}}}}
\newcommand{\uFJ}{{\bm{\mathcal{J}}}}
\newcommand{\uHopf}{\mathcal{H}}

\begin{document}

\title{Three-dimensional magnetization textures as quaternionic functions.}

\author{Konstantin L. Metlov}
\affiliation{Galkin Donetsk Institute for Physics and Engineering, R. Luxembourg str. 72, Donetsk, Russian Federation 283048}
\email{metlov@donfti.ru}

\author{Andrei B. Bogatyr\"ev}
\affiliation{Institute for Numerical Mathematics, Russian Academy of Sciences, 8 Gubkina str., Moscow GSP-1, Russia 119991}
\email{ab.bogatyrev@gmail.com}

\date{\today}

\begin{abstract}
Thanks to the recent progress in bulk full three-dimensional nanoscale magnetization distribution imaging, there is a growing interest to three-dimensional (3D) magnetization textures, promising new high information density spintronic applications. Compared to 1D domain walls or 2D magnetic vortices/skyrmions, they are a much harder challenge to represent, analyze and reason about. Here we build analytical representation for such a textures (with arbitrary number of singularity-free hopfions and singular Bloch point pairs) as products of simple quaternionic functions. It can serve as a language for expressing theoretical models of 3D magnetization textures and specifying a variety of topologically non-trivial initial conditions for micromagnetic simulations. It also follows from the quaternion algebra properties that three dimensional magnetic states can potentially be useful for implementing topological quantum computation.
\end{abstract}

\pacs{75.60.Ch, 75.70.Kw} 
\keywords{micromagnetics, topological solitons, hopfions, Bloch points}

\maketitle

Recent progress in three-dimensional (3D) bulk magnetization vector imaging~\cite{DGSGHRH2017,CMSGHBRHCG2020}, sparked renewed interest to 3D magnetization textures~\cite{Gubbiotti2024roadmap}. Theory of planar domain walls~\cite{MalozemoffSlonczewski} in one dimension is built on top of functions of real variable, representing magnetic moment rotation angle dependence on spatial coordinate. In two dimensions, topology of complex domain walls~\cite{HSG1958} and magnetic vortices/skyrmions in nanostructures can be conveniently described by functions of complex variable~\cite{BP75,M01_CT,M10,BM15_eng,Bogatyrev2017}. Superposition of different topological objects then corresponds to simple products of these functions~\cite{BP75}. In the present work we aim to do this for 3D magnetization configurations. Similar 3D topological textures are observed in liquid crystals~\cite{Chen2013}, colloids~\cite{AS2016}, ferroelectrics~\cite{Lukyanchuk2025} or superfluid $^3He$~\cite{VM1977}, they too may be amenable to such a representation.

Magnets are defined by the existence of their spontaneous magnetization $\vec{M}$, arising in a competition between the quantum-mechanical exchange and thermal fluctuations. The vector $\vec{M}$ is determined at each small (but macroscopic) neighbourhood of a point $\vec{r}$ inside the magnet. Its length $\|\vec{M}\|=\uMs$, called the saturation magnetization, is constant in space, but depends on temperature. It means that statics and dynamics of the magnetization is always a vector rotation. For almost a century, this simple model continues to serve well as a foundation~\cite{LL35} of micromagnetics.

The exchange interaction plays another important role. By assigning a positive energy to spatial variations of magnetization, it smoothes the magnetization vector field $\vec{m}(\vec{r})=\vec{M}(\vec{r})/\uMs$ and makes it a subject of topology, which studies continuity in the most general sense. There are, of course, other energy terms in micromagnetic Hamiltonian (magnetostatic interaction and various forms of magnetic anisotropy~\cite{AharoniBook}, chiral interactions~\cite{Dzyaloshinkskyi58}), which deform the magnetization texture or make some of its configurations more energetically favorable than the others. They are expressed as smooth functions of the magnetization and its derivatives. Away from the phase transition boundaries between configurations of different type, they induce a smooth deformation of the magnetization vector field --- homotopy.

Therefore, our approach here will be to derive the magnetization configurations up to a homotopy, while introducing free functions into them for minimizing the total micromagnetic energy with all the terms, relevant to a particular problem at hand. In a particular case, the present consideration reduces to the model~\cite{M2025} of a magnetic hopfion in helimagnet and can be regarded as a generalization of this previous work.

Just like the complex calculus turned out to be a natural language for describing planar magnetization textures~\cite{M10,Bogatyrev2017}, here we shall adopt the language of quaternions, which is an extension of the complex algebra to higher dimensions. In the text, bold letters will designate quaternion variables $\quat{q}=\quatcomp{w}{a}{b}{c}$. Non-commutative quaternion multiplication, denoted by the dot symbol, follows from the Hamilton's identities: $\uqI\cdot\uqI=\uqJ\cdot\uqJ=\uqK\cdot\uqK=\uqI\cdot\uqJ\cdot\uqK=-1$. The arrows mark purely imaginary quaternions $\vec{r}=\quatcompvec{\ux}{\uy}{\uz}$ (with $w=0$), which are equivalent to 3D vectors. The spatial coordinates here are normalized by a global scale factor $R$ and are considered dimensionless. The normalized magnetization is a vector quaternionic function of a vector quaternionic variable $\vec{m}(\vec{r})$, such that the norm  $\|\vec{m}\|^2=\|\quatcompvec{\umx}{\umy}{\umz}\|^2=\umx^2+\umy^2+\umz^2=1$.

Let us set the boundary condition
\begin{equation}
\label{eq:bndry}
\lim_{\|\vec{r}\|\rightarrow\infty}\vec{m}(\vec{r}) = \uqK,
\end{equation}
so that in the Cartesian coordinate system we choose, the magnetization vector at infinity is aligned with the $\uO\uZ$ axis. The particular direction here is not important, it is only essential that the magnetization at infinity is the same independently on the direction we arrive from. In other words, there is a single infinitely distant point in the space $\vec{r}$. Such a space is called the extended Euclidean $E^3$ and maps to the (three-dimensional) surface of a sphere $S^3$, which itself exists in four-dimensional space. The fixed length magnetization vector endpoints span the surface of a sphere $S^2$, which is the usual sphere in three-dimensional space. The relationship $\vec{m}(\vec{r})$, thus, represents a mapping $S^3\rightarrow S^2$.

{\bf Hopfions.} Let us start with smooth mappings $S^3\rightarrow S^2$. Extending the example of Hopf~\cite{Hopf1931}, Whitehead~\cite{whitehead1947} have shown that all such maps split into integer-numbered homotopy classes. The class number $\uHopf$ is essentially the number of topological solitons (hopfions~\cite{DI79}) in the system. Whitehead's ansatz~\cite{whitehead1947}, based on bi-complex coordinates on the sphere $S^3$, remained the main analytical tool for study of hopfions. It describes vector fields (of an arbitrary Hopf index $\uHopf$) with a central axis. Below we build a quaternionic ansatz for a hopfion ensemble without imposing any global symmetry.

The idea is to build the $S^3\rightarrow S^3$ map and then project the target $S^3$ sphere onto $S^2$. First, note that any unit quaternion $\|\quat{q}\|^2=w^2+a^2+b^2+c^2=1$ represents a point on $S^3$. Alternatively, it can also be represented by a column 2-vector $\{A, B\}$ with two complex numbers, such that $\|A\|^2+\|B\|^2=1$. Elements of $S^3\rightarrow S^3$ map are transforms between two such vectors. In the simplest linear case they can be expressed via unitary $2\times2$ complex matrices from the $U(2)$ group, consisting of the elements of the special unitary group $SU(2)$ with an addition of a complex phase (e.g. multiplying by the matrix $\{\{1,0\},\{0,e^{\imath\eta}\}\}$). The matrices in $SU(2)$ have unit determinants, while the $U(2)$ have $e^{\imath\eta}$. 

The $SU(2)$ matrices directly map to the unit quaternions. Constructing them as a stereographic projection $E^3\rightarrow S^3$ of the vector $\uvr$ and multiplying by a coordinate system axis rotation (around $\uO\uZ$ axis) to represent the additional phase, entering the $U(2)$ matrices, we can write:
\begin{align}
 \label{eq:helicity}
 \uFC(\eta) & = \cos\eta + \uqK\sin\eta\\
 \label{eq:S3S3}
 \uFH(\uvr) & = \uFC(\eta) \cdot \frac{1-\|\uvr\|^2 + 2 \uvr}{1 + \|\uvr\|^2},
\end{align}
where by design $\|\uFH(\uvr)\|=1$.

Unitary quaternions $\quat{U}$, $\|\quat{U}\|=1$ can be used as spinors~\cite{cartan1967} to rotate vector quaternions $\vec{v}^\prime=\quat{U}\cdot\vec{v}\cdot \conj{\quat{U}}$, preserving their length $\|\vec{v}^\prime\|=\|\vec{v}\|$, where the \conjtext{} denotes quaternion conjugation $\conj{\quat{q}}=\conjbr{\quatcomp{w}{a}{b}{c}}=\quatcompconj{w}{a}{b}{c}$. This can be used to restrict $S^3\rightarrow S^3$ mapping~\eqref{eq:S3S3} to $S^2$ subspace, representing the magnetization vector field as:
\begin{align}
\label{eq:Mi}
 \vecsub{m}{i}(\uvr) & = \quat{U}_i(\uvr) \cdot \vecsub{m}{i-1} (\uvr) \cdot \conj{\quat{U}_i(\uvr)}, \\
 \label{eq:Ui}
 \quat{U}_i(\uvr) & = \uFH((\uvr - \uvri)f_i(\|\uvr- \uvri\|)),
\end{align}
where the hopfions, numbered by the index $i$, are added iteratively and $\vecsub{m}{i}(\uvr)$ represents the magnetization vector field at $i$-th iteration, $\uvr_i$ is the location of $i$-th hopfion and $f_i(\|\uvr\|)$ is its profile function (such that $\lim_{x\rightarrow0}x f_i(x) =0$ and $\lim_{x\rightarrow\infty}x f_i(x) =\infty$), which can be used to strongly localize the hopfion in space.

\begin{figure*}[tb]
 \includegraphics[width=\textwidth]{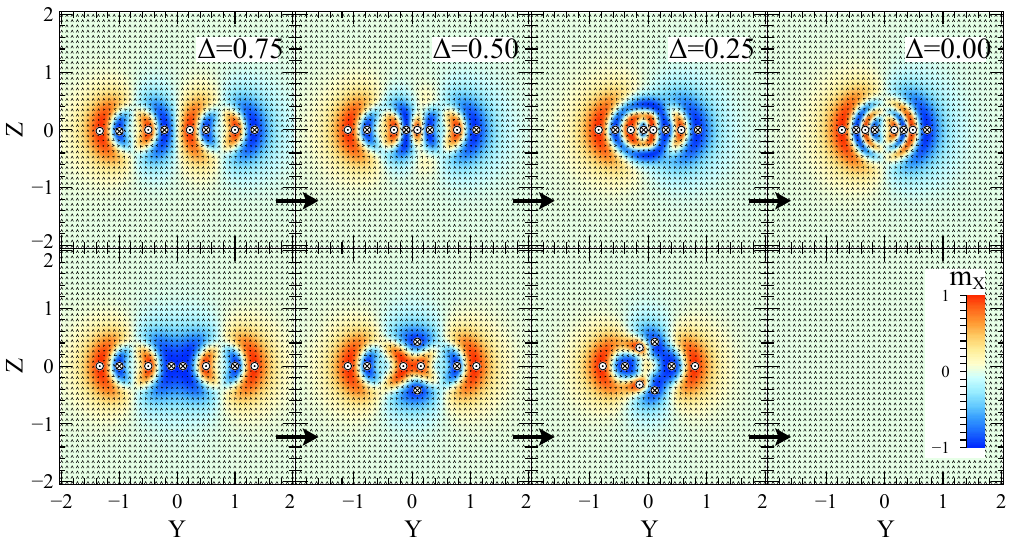}
 \caption{\label{fig:annihilation}Merger (top row) and annihilation (bottom row) of hopfions. The top row configuration is $\quat{U}=\uFH_1(\protect\uvr+\protect\uqJ\Delta)\cdot \uFH_1(\protect\uvr-\protect\uqJ\Delta)$ and the bottom row is $\quat{U}=\uFH_1(\protect\uvr+\protect\uqJ\Delta)\cdot \conj{\uFH_1(\protect\uvr-\protect\uqJ\Delta)}$,
 where $\uFH_1(\protect\uvr)=\uFH(\protect\uvr f(\|\protect\uvr\|))$, $\eta=0$ and $f(x)=6(x+2x^5)$ localizes the isolated hopfion approximately inside a unit sphere. The magnetization vector field is given by~\eqref{eq:allM} with $\protect\vecsub{m}{0}=\protect\uqK$.
 }
\end{figure*}
To prove that~\eqref{eq:Mi} indeed represents a hopfion, we can reduce it to 
the model~\cite{M2025} for which pre-images of the corresponding $S^3\rightarrow S^2$ map are computed in the Appendix~A of~\cite{M2024ms} and have the linking number of $1$. This is done by setting $r_1=0$, $f_1(x)=(1/x) e(x)/(1-e(x))$ when $0<x<1$, with the profile function $e(x)$ such that~\cite{M2025} $e(0)=0$, $e(1)=1$, $f_1(x)=\infty$ when $x>1$, putting $\eta=\pi/4-\chi/2$, and starting with the uniform magnetization $\vecsub{m}{0}=\uqK$. The magnetization vector field of~\cite{M2025} is then reproduced up to the chirality change, described in the Fig.~1 caption of~\cite{M2025}. This change is mere a convenience to make the hopfions~\eqref{eq:S3S3} correspond to helimagnets with positive Dzyaloshinskii-Moriya interaction constant. Ultimately, this magnetization distribution (up to chirality reversal and rescaling) coincides with the original Whitehead's ansatz~\cite{whitehead1947} for $\uHopf=1$ hopfions.

Therefore, one step of the iteration~\eqref{eq:Mi} adds $1$ to the Hopf index of the magnetization distribution. With more steps, arbitrarily complex hopfion configurations can be built. They are represented by a quaternionic product
\begin{equation}
 \label{eq:Uprod}
 \quat{U}(\uvr) = \prod\limits_{i} \quat{U}_i(\uvr)
\end{equation}
and applied to an arbitrary initial magnetization configuration $\vecsub{m}{0}(\uvr)$ all at once
\begin{equation}
\label{eq:allM}
\vec{m}(\uvr) = \quat{U}(\uvr) \cdot
\vecsub{m}{0}(\uvr) \cdot \conj{\quat{U}(\uvr)}.
\end{equation}
Note that in quaternionic algebra $\conjbr{\quat{a}\cdot\quat{b}}=\conj{\quat{b}}\cdot\conj{\quat{a}}$.

It is also interesting that conjugate of a unitary matrix is its own inverse and so is the conjugate of a unit quaternion $\quat{U}\cdot\conj{\quat{U}}=1$. Therefore, if instead of some $\quat{U}_i$ in the product~\eqref{eq:Uprod} we use its conjugate $\conj{\quat{U}_i}$, the Hopf index will {\em decrease} by $1$ on this iteration step. This means that conjugate of a hopfion is anti-hopfion. In fact, any complex hopfion configuration $\quat{U}$ can be turned into the corresponding anti-hopfion by conjugating it.

The existence of the $\uHopf=\pm1$ hopfions and anti-hopfions was already established in~\cite{Guslienko2025} on the basis of rotation matrix analysis, but the present approach allows to represent the whole process of the hopfion merger and annihilation for an arbitrary Hopf index. The top row in Fig.~\ref{fig:annihilation} shows the merger of two identical $\uHopf=1$ hopfions into a single $\uHopf=2$ hopfion, while the bottom row shows the annihilation of the $\uHopf=1$ hopfion with the corresponding anti-hopfion into the uniformly magnetized state. The plot is only able to capture the hopfion cross section, as the whole process takes place in 3D. But a cross-   section of a continuous 3D vector field is a continuous 2D vector field and therefore its continuous transformation must preserve 2D topological charge. One can see that in the cross section, shown in Fig.~\ref{fig:annihilation} top row, the annihilation proceeds through merger of vortex and antivortex filaments with co-aligned out-of-plane core magnetizations (relatively to the cross-section plane). Such pairs in 2D are topologically trivial~\cite{Tretiakov2007} and their unwinding can happen smoothly. The whole process in Fig.~\ref{fig:annihilation} is a homotopy --- a continuous deformation of the magnetization vector field as function of the parameter $\Delta$. The topological index is therefore independent of $\Delta$. It is equal to $2$ in all frames of the top row and is equal to $0$ in all frames of the bottom row. One can also observe from $\Delta=0.75$ frames that two hopfions are visibly repelled from each other, while the hopfion and the anti-hopfion are immediately attracted to each other.

Another interesting feature in Fig.~\ref{fig:annihilation} is asymmetry with respect to $\uY=0$ plane. Merger and annihilation do not just happen to both hopfions, but one hopfion always consumes the other. This is due to non-commutativity of quaternion multiplication. Should the order of hopfions in the product $\quat{U}$ in Fig.~\ref{fig:annihilation} be reversed, the frames will become mirrored across $\uY=0$ plane, but the asymmetry will remain. Multiplication of unitary matrices, describing the $S^3\rightarrow S^3$ mapping, is also non-commutative. This implies that hopfion merger and annihilation are inherently accompanied by symmetry breaking. 
It follows here from the pure geometry, but there is also a physical reason for it. Owing to the Hobart-Derrick theorem~\cite{Hobart1963,*Derrick1964} the classical chirally-symmetric micromagnetic Hamiltonian can not support statically stable magnetic hopfions. The lowest order energy terms in the Landau order parameter expansion, enabling their existence, correspond to the Dzyaloshinskii-Moriya antisymmetric exchange, which breaks the chiral symmetry between the mirror images of the same system.

{\bf Bloch points (BPs).} There is another wide class of 3D magnetization patterns, which, unlike hopfions, contain {\em singularities} of the magnetization vector field~\cite{Feldtkeller1965}. In lower dimensions the field can avoid singularities by escaping into a higher dimension. A prime example of such a phenomenon is the magnetic vortex core~\cite{UP93}. But, in three dimensions such escape is not possible and thus a theory of three dimensional magnetization patterns must include singular configurations as well. Fortunately, they too map into products of quaternionic functions. But, this time of normalized to unit modulus (everywhere, except a discrete set of singular points) vector quaternions.

\begin{figure}[tb]
 \includegraphics[width=\columnwidth]{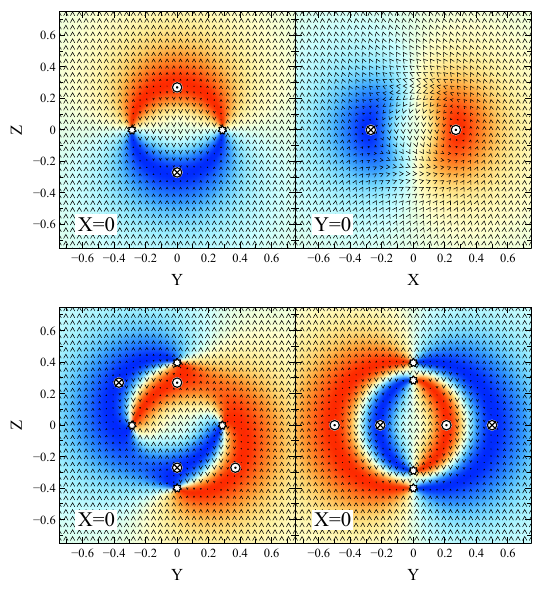}
 \caption{\label{fig:BPs}Various Bloch point configurations. Top row shows two different cross sections of a BP pair, $\vec{m}(\uvr)=\uFC\cdot\uFR\cdot\uFB(\uvr f(\|\uvr\|),-\uqJ/2,\uqJ/2)\cdot\conj{\uFR}\cdot\conj{\uFC}$ with $\uFC=\uFC(\pi/4)$, $\uFR=\uFR(\uqJ,\uqK)$, and $f(x)$ like in Fig.~\ref{fig:annihilation}. The bottom row shows combinations $\vec{m}(\uvr)=\uFC\cdot\uFR\cdot\uFJ(\uFB_1,-\vecsub{b}{1},\vecsub{n}{1},\uFB_2,-\vecsub{b}{2},\vecsub{n}{2})\cdot\conj{\uFR}\cdot\conj{\uFC}$, $\uFB_1=\uFB(\uvr f(\|\uvr\|),-\vecsub{b}{1},\vecsub{b}{1})$, $\vecsub{n}{1}=\vec{n}(\uvr,-\vecsub{b}{1},\vecsub{b}{1})$, $\uFB_2=\uFB(\uvr f(\|\uvr\|),-\vecsub{b}{2},\vecsub{b}{2})$, $\vecsub{n}{2}=\vec{n}(\uvr,-\vecsub{b}{2},\vecsub{b}{2})$ with (left) $\vecsub{b}{1}=\uqJ/2$, $\vecsub{b}{2}=-\uqK$ and (right) $\vecsub{b}{1}=-\uqK$, $\vecsub{b}{2}=\uqK/2$. Stars mark positions of point singularities.
 }
\end{figure}
If we limit ourselves to localized in space magnetization configurations, it is still necessary to match the boundary condition~\eqref{eq:bndry}. This is only possible if singularities are added in pairs: a point, described by a unit-length vector quaternion function $\vec{B}(\vec{r}-\vecsub{b}{1})$, and the corresponding anti-point, described by $\conj{\vec{B}(\vec{r}-\vecsub{b}{2})}$, where $\vec{B}(\vec{r})=\vec{r}/\|\vec{r}\|$. Quaternionic product of two such functions has unit norm. It unwinds when positions of the point $\vecsub{b}{1}$ and the anti-point $\vecsub{b}{2}$ coincide, otherwise the magnetization on a small sphere around each of the singularities has a unit 2D topological charge~\cite{doring68,VM1977liqcr}. The whole distribution can be represented via the function $\uFB (\vec{r},\vecsub{b}{1},\vecsub{b}{2})$
\begin{align}
 \label{eq:BB}
 \uFB & =
 \frac{\vecsub{b}{1}-\vecsub{b}{2}}{\|\vecsub{b}{1}\!-\!\vecsub{b}{2}\|}\cdot\frac{\vec{r} - \vecsub{b}{1}}{\|\vec{r}\!-\!\vecsub{b}{1}\|}\cdot\conjwbs{\frac{\vec{r} - \vecsub{b}{2}}{\|\vec{r}\!-\!\vecsub{b}{2}\|}}, \\
 \label{eq:MBB}
 \vecsub{m}{\mathcal{B}} & =
 \uFB(\vec{r}f(\|\vec{r}\|),\vecsub{b}{1},\vecsub{b}{2}),
\end{align}
where the $f(x)$ controls the degree of localization. This by design represents a unit length vector $\vecsub{m}{\mathcal{B}}$ and has a pair of singularities. Why it is not enough to consider just a product of two last terms in~\eqref{eq:BB}? The reason is that the product of two vector quaternions may also have non-zero real part, proportional to the scalar product of the vectors. But the triple product~\eqref{eq:BB} is guaranteed to be a vector since all three vector quaternions (and, consequently, their product) lie in the same plane.

To match the boundary condition~\eqref{eq:bndry} exactly, we need to rotate the magnetization in such a way that the direction $\vecsub{b}{1}-\vecsub{b}{2}$ becomes $\uqK$. Such rotation can be done with a unitary quaternion
\begin{equation}
 \label{eq:rotation}
 {\uFR(\vecsub{d}{1},\vecsub{d}{2})} = 
 \frac{\sqrt{\vecsub{d}{2}\cdot\conj{\vecsub{d}{1}}}}{\sqrt{\|\vecsub{d}{1}\|\|\vecsub{d}{2}\|}}.
\end{equation}
A special care is needed to avoid indeterminate result when the vectors are exactly opposite ($\vecsub{d}{2}=-\vecsub{d}{1}$). This case is just a reflection or rotation by $\pi$ around {\em any} axis in the plane, perpendicular to both vectors

The example of a single Bloch point pair~\eqref{eq:BB} with rotation~\eqref{eq:rotation} and an additional helicity~\eqref{eq:helicity} is shown in Fig.~\ref{fig:BPs} top row. One can see both the plane, cutting through the singularities (left), and the plane in the middle between them (right), where a vortex-antivortex pair is clearly visible. Such vortex and antivortex filaments often run between the Bloch points of the opposite charge~\cite{CMSGHBRHCG2020}.

These BP pairs can be combined into more complex configurations using the same approach with triple quaternionic product. The trick is to rotate the planes of each combined BP pair in such a way that they coincide and chose a pre-factor vector $\vec{v}$, lying in the same plane. The normal to a single BP pair plane is aligned with a vector product $\vec{n}(\uvr, \vecsub{b}{1}, \vecsub{b}{2})$
\begin{equation}
 \vec{n}=\frac{(\vec{r}-\vecsub{b}{1})\cdot(\vec{r}-\vecsub{b}{2}) - (\vec{r}-\vecsub{b}{2})\cdot(\vec{r}-\vecsub{b}{1})}{2},
\end{equation}
and its magnetization at infinity is $\vec{m}^\infty=\vec{B}(\vecsub{b}{1}-\vecsub{b}{2})$.

Suppose we are combining BP pairs with magnetizations $\vecsub{m}{1}(\uvr)$, $\vecsub{m}{2}(\uvr)$, normals $\vecsub{n}{1} (\uvr)$, $\vecsub{n}{2}(\uvr)$ and magnetizations at infinity $\vec{m}^\infty_1$, $\vec{m}^\infty_2$. First, let's select a new normal $\vec{n}=\vec{B}(\vecsub{n}{1}+\vecsub{n}{2})$ in the middle between the original two (when $\vecsub{n}{1}=-\vecsub{n}{2}$ the normals essentially coincide and nothing needs to be rotated). Then we introduce rotations of the magnetization of the BPs towards this common normal $\uFR_1 = \uFR(\vecsub{n}{1},\vec{n})$ and $\uFR_2 = \uFR(\vecsub{n}{1},\vec{n})$. For the triple product we can, basically, select any vector $\vec{v}(\vec{n},\beta)$ in the plane, normal to $\vec{n}$, which can be parametrized by a rotation angle $0\le\beta<2\pi$. Finally, the function, combining the two BP pairs is
$\uFJ(\vecsub{m}{1},\vec{m}^\infty_1,\vecsub{n}{1},\vecsub{m}{2},\vec{m}^\infty_2,\vecsub{n}{2},\beta)$
\begin{align}
 \uFJ &=\frac{\vec{v}}{\|\vec{v}\|}\cdot
 \uFR_1\cdot\vecsub{m}{1}\cdot\conj{\uFR_1}\cdot
 \uFR_2 \cdot \vecsub{m}{2} \cdot \conj{\uFR_2},
\end{align}
where $\vec{v}=\vec{n}\cdot\vec{\mu} - \vec{\mu}\cdot\vec{n}$ is used and $\vec{\mu}=\uFR_1\cdot\vec{m}^\infty_1\cdot\conj{\uFR_1}$.

When only two BP pairs are combined, the choice of $\beta$ impacts only the direction of the magnetization at infinity, which is restored by performing the final rotation. But if more BPs are combined, the value of $\beta$ influences the direction of the domains, forming in between the Bloch points and is another degree of freedom. Also, the freedom to rotate the magnetization around the $\uO\uZ$ axis using the unit quaternion $\uFC(\xi)$ applies to BPs as well as to hopfions. The examples of combined BP pairs are shown in the bottom row in Fig.~\ref{fig:BPs}.

Arbitrarily more complex configurations of BPs and hopfions can be built using the attached Mathematica code~\cite{supplemental3D}, implementing the above expressions. In all cases 3D magnetization vector fields are expressed as products of simple quaternionic functions. This basic language is amenable to many further generalizations. But even in the present form it can already generate many useful trial functions for analytical modeling (e.g. of magnetic globules~\cite{Beg2019}, half-globules like chiral bobbers~\cite{RBBK2015}) and provide analytical initial configurations for numerical micromagnetics.

Besides the considered 3D configurations, 2D and 1D magnetization distributions (such as cones, helices or skyrmions) can also be expressed as quaternionic functions. This allows to combine them with hopfions using~\eqref{eq:allM} to model e.g. heliknotons~\cite{Kuchkin2023}, which, however, is beyond the scope of the present paper.

There is also a very general consequence, implied by the use of quaternionic products. Such algebra of 3D magnetic states is inherently non-Abelian. It describes not only the field of interacting objects (solitons), but implies that the field also contains the imprint of the order in which the solitons were brought into the interaction. It is already known that non-Abelianness of the quaternionic group results in nontrivial rules of coalescence of disclinations in cholesteric liquid crystals~\cite{VM1977liqcr}. In magnetism it can lead to many interesting properties of three-dimensional topological solitons as well. In particular, non-Abelian algebra of multi-soliton states implies the possibility to implement a topological quantum computation according to Kitaev's scenario~\cite{Kitaev2003}, once the methods of nano-scale nucleation and control of these three-dimensional states are perfected.

Support of the Russian Science Foundation under the project {\tt 25-11-00144} is gratefully acknowledged.

%
\end{document}